\documentclass[useAMS,usenatbib,usegraphicx]{mn2e}
\usepackage{amsmath}
\title[Reverberation in 1H0707-495]{Understanding Reverberation Lags in 1H0707-495}
\author[Zoghbi et. al]{A. Zoghbi$^{1}$\thanks{E-mail:
azoghbi@ast.cam.ac.uk}, P. Uttley$^{2}$ and A. C. Fabian$^{1}$\\
$^{1}$Institute of Astronomy, Madingley Road, Cambridge CB3 0HA\\
$^{2}$School of Physics and Astronomy, University of Southampton, Highfield, Southampton SO17 1BJ}
\begin{document}

\date{}

\pagerange{\pageref{firstpage}--\pageref{lastpage}} \pubyear{2010}

\maketitle

\label{firstpage}

\begin{abstract}
The first reverberation lag from the vicinity of a supermassive black
hole was recently detected in the NLS1 galaxy 1H0707-495.  We
interpreted the lag as being due to reflection from matter close to the black
hole, within a few gravitational radii of the event horizon (an inner
reflector).  It has since been claimed by Miller et al that the lag
can be produced by more distant matter, at hundreds of gravitational
radii (an outer reflector). Here, we critically explore their
interpretation of the lag. The detailed energy dependence of the time
lags between soft and hard energy bands is well modelled by an inner
reflector using our previously published spectral model. A contrary
claim by Miller et al was obtained by neglecting the blackbody
component in the soft band.  Soft lags can be produced by a
large-scale outer reflector if several, implausible, conditions are
met.  An additional transfer function is required in the soft band corresponding
to a region that is physically close to the continuum source, or lies close to our line of sight
and subtends a small
solid angle at the source, challenging the production of the observed spectrum.  We show that the
original
inner reflector interpretation of reverberation very close to the
black hole provides a self-consistent and robust model which explains
the energy spectrum and timing properties, including the time delays,
power spectra and the shape of the coherence function. Several of
these properties are opposite to the predictions from a simple
large-scale outer reflection model.
\end{abstract}


\section{Introduction}
X-ray emission from black holes is known to be highly variable on
short time scales, down to milliseconds in Galactic black holes, and
to minutes in active galactic nuclei (\citealt{1999ApJ...510..874N,
2000A&A...363.1013R,2004MNRAS.348..783M}). The emission region must therefore be 
compact and close to the black hole. Further evidence is obtained from the X-ray spectra. 
 The first relativistically
broadened iron K line was detected in ASCA data
(\citealt{1995Natur.375..659T}), and later confirmed with XMM-Newton
(\citealt{2002MNRAS.335L...1F}) and in many other AGN (e.g.
\citealt{2007MNRAS.382..194N}, see \citealt{2007ARA&A..45..441M} for a review). Such broad iron
lines are also seen in Galactic black holes
(\citealt{2007ARA&A..45..441M,2010MNRAS.402..836R}) and neutron stars
(\citealt{2008ApJ...674..415C}).

More recently, a broad iron L line at energies $<1$ keV has  been detected
(\citealt{2009Natur.459..540F,2010MNRAS.401.2419Z}, hereafter F09 and Z10 respectively) in the
Narrow Line Seyfert 1 (NLS1) galaxy 1H0707-495 ($z = 0.0411$). The line is visible thanks to the
high reflection fraction and the relatively high iron abundance. Modelling the line as part of a
reflection continuum, smeared by relativistic effects, indicates that spectrum is emitted from a
very small region (less than $\sim 3$ gravitational radii, $r_g = GM/c^2$, F09,Z10).

Along with this identification, came the detection of a time delay of $\sim 30$ seconds between
the direct and reflected emission. It was measured between the 0.3--1.0 and 1.0--4.0~keV bands, the
first being dominated by reflection (iron L line) and the second dominated by the direct power-law
emission. This time lag was interpreted as a reverberation signature from the vicinity of the black
hole (F09, Z10), i.e. from a small-scale reflector with a size-scale of tens of light-seconds,
namely the inner disk which produces the strong reflection signatures in this source.

The lag spectrum of 1H0707-495 shows these reverberation delays (soft lags) over a broad range of
high temporal
frequencies ($> 5\times10^{-4}$ Hz), while at low frequencies, it shows hard lags similar to
those commonly seen in GBH and other AGN
(\citealt{1989Natur.342..773M,1999ApJ...510..874N,2000A&A...357L..17P,2007MNRAS.382..985M,
2008MNRAS.388..211A}). The exact origin of these hard lags is not clear (see for example
\citealt{2001AdSpR..28..267P}). Comptonisation can possibly produce such a delay, but a picture in
which soft and hard photons are emitted from slightly different regions, used within a model of
fluctuations propagating through the accretion flow, is more consistent with other variability
properties
(\citealt{1997MNRAS.292..679L,2001MNRAS.327..799K,2006MNRAS.367..801A}).

\cite{2010MNRAS.403..196M} (hereafter M10a) put forward a model in which these hard lags are due to
reflection from distant material (few hundreds to few thousands
gravitational radii). This model was able to reproduce the hard lags
in NGC 4051 (M10a), previously explained with the propagating
fluctuations model (\citealt{2004MNRAS.348..783M}). In
\cite{2010arXiv1006.5035M} (hereafter M10b), the same model was
applied to 1H0707-495 claiming that it can produce soft lags. In the
model, the sharp edges in the transfer function of the new distant
reflector produce oscillations that can lead to small negative lags
(i.e soft lags in this case).  These oscillations lead the authors to
caution against interpreting the observed lags as being due to light
travel time due to reverberation with an intrinsically short light
travel time from the primary continuum source to the reflector.
Instead, they suggest that the lags originate from a large-scale
reflector, such as a disk-wind
(e.g. \citealt{2010MNRAS.404.1369S}). No spectral model is presented
in M10b.

In the present work, we critically compare the models for large-scale,
outer, and small-scale, inner, reflection which are used to explain
the lags in 1H0707-495 by M10b and Z10 respectively.  We use the
energy-dependence of the lag to show that, contrary to a claim by
M10b, a small-scale reflector fits the data well, especially when we
take into account lags from the thermally-reprocessed blackbody
emission which is seen in the energy spectrum.  We also show how the
models fit the Fourier frequency-dependence of the lags and highlight
the fact that both models {\it require} reflectors with a small
time-delay, which can be naturally explained if the reflector is
small, but leads to a constrained geometry, which is unfavourable to the required large covering
fraction for the large-scale outer reflection model of M10b.

\section{Modelling the lag}\label{lag_modelling}
\subsection{Lag vs Energy}\label{energy_dependence}
In this section, we explore the energy dependence of the lag.
The data presented in this section are from previous XMM-{\it Newton} observation of 1H0707-495
in 2002, 2007 and 2008. The details of the reduction are similar to that in Z10.
Background-subtracted light curves in ten energy bands were produced
(defined by equal distances in log-space over two broad bands). For each energy band,
frequency-dependent time lags were
calculated between the light curve in that band and a light curve of the whole energy
band (0.3--10 keV) excluding the current band. Using the whole band as a reference maximises the
signal to noise ratio, while excluding the current band from the reference ensures the noise remains
uncorrelated between the bands. This however means the reference light curve is slightly different
for each band, but this has very minor effect on the final results. Monte Carlo simulations showed
that there is about 1 percent systematic error in the lags measured this way which is much smaller
than the statistical errors of the current data. Using one of the bands as a reference (the softest
band as it has the highest number of counts) gives the same result but with slightly larger error
bars.

\begin{figure}
\centering
 \includegraphics[width=220pt,clip ]{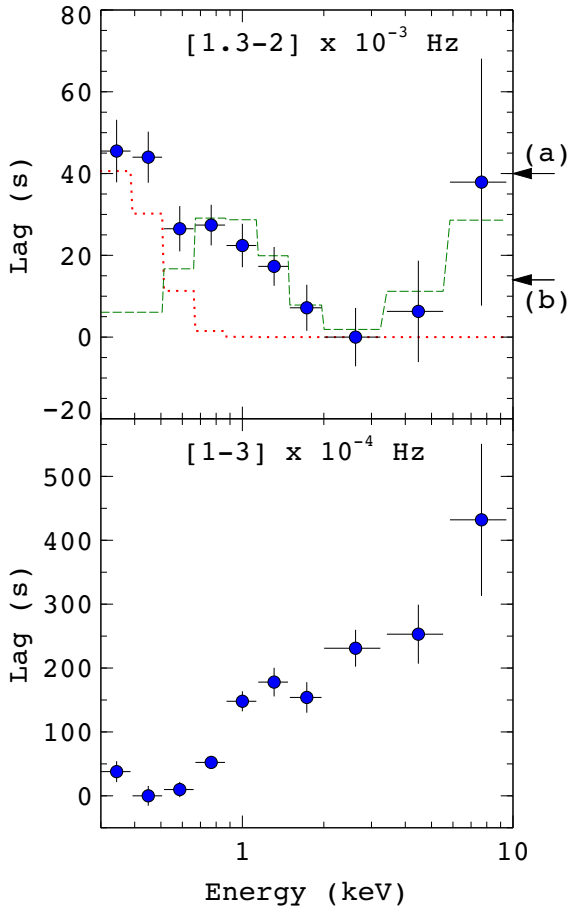}
\caption{\emph{Top:} Energy-dependence of the lag at $[1.3-2]\times10^{-3}$ Hz. The lag is
calculated with respect to the light curve in the whole band (see text), then shifted so that the
minimum corresponds to zero. The two arrows on the right mark the average lag if wider energy bins
are considered, {\tt(a)} is for 0.3--1 keV and {\tt(b)} is for 4--7.5 keV. The green dashed line is
a simple model representing the ratio of reflection to the whole spectrum (plus a constant offset
for the reference band and with a multiplying constant to represent the intrinsic lag). The red
dotted line represents the lag model attributed to the blackbody component. \emph{Bottom:} The same
plot for the
$[1-3]\times10^{-4}$ Hz band.}
\label{fig:lag_vs_energy}
\end{figure}

The top panel of Fig. \ref{fig:lag_vs_energy} shows the
variation of the lag as a function of energy in the $[1.3-2]\times10^{-3}$ Hz frequency band, these
are the frequencies where the soft lag is at a maximum (see Fig. 18 in Z10).  The choice of
reference band sets a constant offset to the time-lag, but what is important are the differences in
lag between energy bands: more positive values lag smaller/more-negative values.  
The lag is maximised between the soft band and energies of $\sim3$~keV where the primary continuum
dominates the spectrum. This shape is strikingly
similar to the (inverted) shape of the fractional RMS variability spectrum presented in
F09, and is different from the positive log-linear correlation of
lag with energy seen in galactic black holes and other AGN (\citealt{1989Natur.342..773M}).

\begin{figure}
\centering
 \includegraphics[width=240pt,clip ]{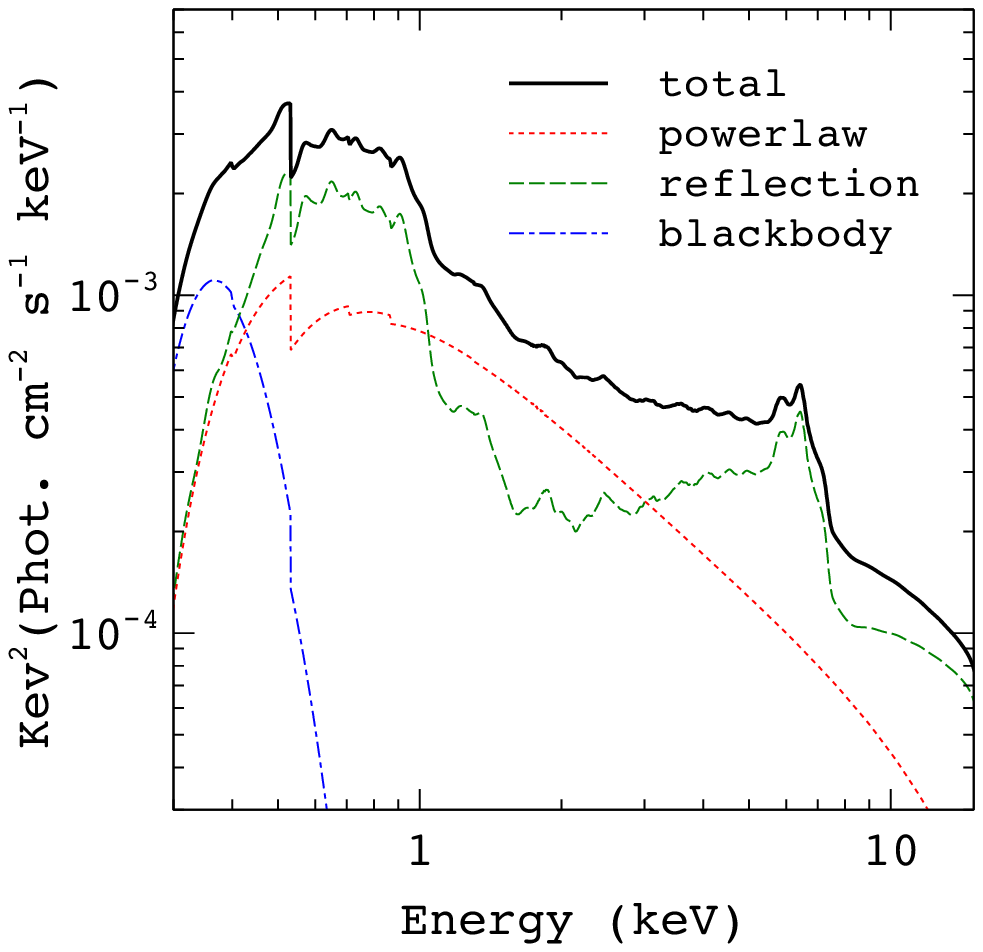}
\caption{The best fitting model to the energy spectra described in Z10. The plot is in the form
$E^2 f(E)$. It shows the main components of spectrum. The reflection component is a small-scale
reflector that is blurred by relativisitc effects.}
\label{fig:plot_eem}
\end{figure}

This shape can be easily understood in light of a model where soft
lags are reverberation signatures from very close to the black
hole. In a similar way to the interpretation of the rms spectrum, the
lag is maximised between bands which are dominated respectively by
reflection and direct emission (i.e. the power-law in
Fig. \ref{fig:plot_eem}). The lag at high energies (dominated by the
broad iron K line) has similar values to those of the iron L line ($<
1$ keV), which is expected if they are indeed representing the same
reflection components. There appears also to be a positive hard lag
between the 0.3--0.5 keV band and the 0.5--1 keV band.  The second
band is dominated by the relativistically smeared reflection, while
the first is mainly blackbody emission (Fig. \ref{fig:plot_eem}).  It
is natural to interpret the 0.3-0.5 keV lag as being associated with
thermal reprocessing of X-ray photons absorbed by the disk, which is
expected to accompany reflection from the hot inner disk in an
extreme-accretion-rate AGN.

A simple model based on this interpretation is also shown in Fig. \ref{fig:lag_vs_energy}. This
model is given by the intrinsic average time delay from the direct continuum to the reprocessed
component, multiplied by the ratio of the reprocessed emission (reflection or thermally reprocessed
blackbody emission) to the total emission including the
power-law, both taken from the best fit model (Fig. \ref{fig:plot_eem}) described in detail in Z10.
The model naturally produces the lag between hard and soft bands and the medium-energy band where
the direct power-law contribution to emission is maximised.  The intrinsic lags that give the best
combined fit are $70\pm20$~s for the reflector and $110\pm10$~s for the thermally-reprocessed
blackbody emission, corresponding to a few gravitational radii for a black hole of mass
$\sim10^7$~M$_{\odot}$.  We
expect there to be a small difference between the light-travel
times to
the reflector and the thermal reprocessor, because where reflection is maximised due to disk
ionisation, thermal reprocessing of absorbed flux should be minimised and vice versa.  Hence a
radial dependence of ionisation could easily explain the small difference in the light-travel time
to these components. The actual values of the lag are different from those in Fig.
\ref{fig:lag_vs_energy} because the energy bands include contribution from both different
components whose lag is diluted, while the model fits for each individual component.

M10b calculated the lag between 0.3--1 and 4--7.5 keV and found that there is a soft delay in the
$[1-2]\times10^{-3}$ Hz frequency band. This led the authors to conclude that such a significant
lag between the L-line and the K-line reflection parts of the spectrum rules out the small-scale
reflector interpretation,
which requires that K and L lines share the same location in the inner disk. It is apparent from
Fig.
\ref{fig:lag_vs_energy} that the lag at the
iron K energies is actually consistent with that of the L-line present just below 1 keV. The energy
bands chosen in M10b are broad and given the steepness of the energy spectrum, the
measured lag has a significant contribution from energies below 0.5 keV, which in our spectral model
is dominated by blackbody emission and \emph{not} reflection. This can be seen in Fig.
\ref{fig:lag_vs_energy} where the average lags at the 0.3--1 and 4.--7.5 keV bands used in M10b are
marked with arrows on the top-right of
Fig. \ref{fig:lag_vs_energy}. The difference between the two is the
lag that would be calculated when those two bands are considered, and
shows a clear lag of the hard relative to the soft band, as reported
by M10b.  However, the detailed lag vs. energy spectrum in
Fig. \ref{fig:lag_vs_energy} shows that this result is expected due to
the domination of the 4-7.5 keV band by the lower-energy photons in
the band, which leads to a stronger direct continuum contribution,
together with the blackbody component dominating the softest
energies. M10b did not take account of the blackbody component in
their comparison with the expectations of the small-scale reflector
model.

The bottom panel of Fig. \ref{fig:lag_vs_energy} is similar to the top
panel but now showing the lags in the $[1-3]\times10^{-4}$ Hz frequency
band. This, unlike the top panel, shows a lag that increases with
energy, similar to that commonly seen in many accreting sources, both
Galactic binaries and AGN
(\citealt{1989Natur.342..773M,1999ApJ...517..355N,2008MNRAS.388..211A}). The
energy dependence of the lag in the two frequency bands
($[1-3]\times10^{-4}$, bottom and $[1.3-2]\times10^{-3}$~Hz, top in
Fig.~\ref{fig:lag_vs_energy}) is not simply an inverted version of the
top panel (as would be expected if the soft lags are simply an
artefact due to the sharp edge in the transfer function of a large-scale reflector) 
suggesting that two \emph{separate} components are contributing to the
lag spectrum. As we pointed out in F10 and Z10, all the evidence
points towards the lag at $[1-2]\times10^{-3}$ Hz being due to
reflection from a distance of a few gravitational radii. The
energy-dependence of the lags at low frequencies shows no clear link
with the reflection spectrum (except possibly for the very low
energies where the blackbody dominates). This leads us to postulate
that the long time scale lags may be driven by fluctuations in the
accretion flow which modulate the blackbody disk emission before
reaching the direct power-law emitting region to produce the observed
long-time-scale lags (e.g.
\citealt{2006MNRAS.367..801A}).

\subsection{Lag vs Frequency}\label{dist_ref}
In this section we explore the lag dependence on frequency and how it
can be explained using models of a small-scale and large-scale,
i.e. inner and outer reflector.

The model of a large-scale reflector presented by M10a and expanded in
M10b explains the low-frequency hard lags in terms of light-travel
time to a distant reflector with a significant contribution in the
hard band (note that this is the opposite of the inner-reflection model, where reflection is
stronger in the soft band). This produces a constant positive hard lag at low
frequencies, which breaks at a time scale comparable to the light
travel time at the outer radius of the scattering medium. In this
model, lag transfer functions with sharp edges (e.g. a top hat
function) produce oscillations in the lag spectra. If small lags from
close to the line of sight are removed, these oscillations may go
negative at high frequencies.  However, as shown by M10b, it is
difficult to obtain negative lags that are as broad in frequency as those
seen in the data. There are two main reasons for this. First, the
oscillations in lag have frequency-width of order $1/\tau_0$, where
$\tau_0$ is the largest lag produced in the transfer function.  The
magnitude of the negative lag itself scales with $\tau_0$. As a
consequence, making the lag more negative forces it to be
narrower. The second reason is the fast decay of the response of the
hard reflector at these short time scales. Fast variability is smeared
out by the extended transfer function, and the lag tends to zero as
the source variability is now dominated by the direct emission.
\begin{figure}
\centering
 \includegraphics[height=220pt,clip ]{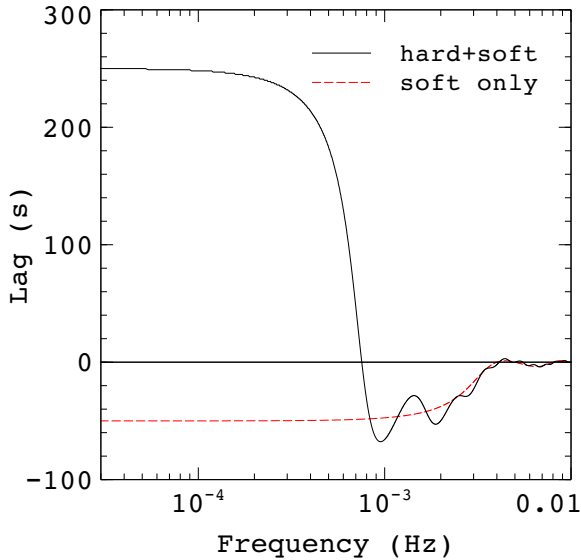}
\caption{A distant reflector model with two constant transfer functions. They extend between
200--1000 s in the hard band and 0--200 s in the soft band. The continuous black line shows the lag
when both components are included. The dashed red line is when the transfer function in the hard
band is set to zero (soft only). The plot shows that the broad negative trough in the hard+soft case
is mainly produced by the soft transfer function.}
\label{fig:decomposed}
\end{figure}

This suppression of reflected emission at high frequencies was overcome in M10b by introducing an
additional top-hat transfer function in the soft band, which cuts off at a much shorter time-lag
(150~s).  The effect of this component is to reproduce the broad negative lag above the hard-lag
cut-off in the lag spectrum, as demonstrated in Fig. \ref{fig:decomposed}, which shows the
contributions of the transfer functions in the soft and hard bands.  The soft reflection component,
due to its narrow transfer-function, dominates the lag at high frequencies, up to frequencies
corresponding to the maximum lag in the transfer function where the reflector variability becomes
suppressed.  At high frequencies, the effect of the hard lagging component is only to produce the
narrow oscillations superimposed on the lags, which are not detectable in the data.  Therefore, even
for the large-scale reflection model, where the low-frequency hard lags are produced by reflection,
a soft band reflector is required which is of much smaller scale, at least in terms of the width of
the transfer function.  M10b suggest that the small lags of this component need not correspond to a
reflector of small intrinsic size, since the lags represent the light-travel delay between the
direct continuum and the reflected signal seen by the observer, which depends on the source and
reflector geometry and does not simply correspond to the actual light-travel time between the
continuum source and the reflector.  However, as we will argue in Section~\ref{discussion}, a
constraining geometry is required in order for the light-travel time from the continuum
to the reflector to be very different from the observed delay.
\begin{figure}
\centering
 \includegraphics[height=200pt,clip ]{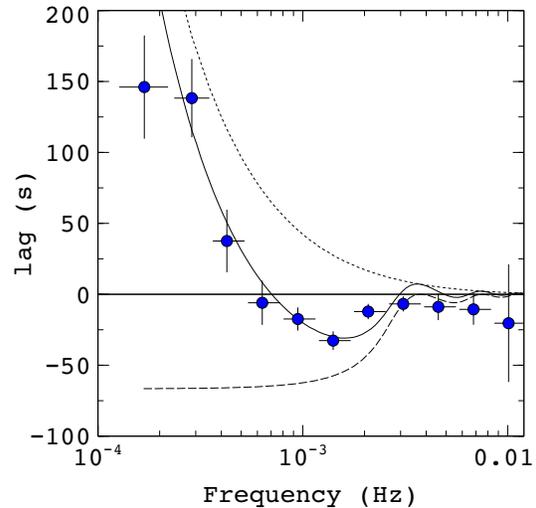}
\caption{The lag spectrum of 1H0707-495 (data points), along with a two component model. The dotted
line is a model of propagating fluctuations (see \citealt{2006MNRAS.367..801A} for details), the soft
lag is modelled here by a simple reflector with a uniform transfer function. The continuous line is
the total lag taken as
the sum of the two components.}
\label{fig:fluc_model}
\end{figure}

The small-scale reflection model of F09 and Z10 requires a separate,
non-reflection origin for the large, low-frequency hard lags seen in
the data.  This interpretation is supported by the fact that, as shown
in Sec. \ref{energy_dependence}, the lags at low and high frequencies
appear to be spectrally-independent.  It has been shown that the
propagating fluctuations model can explain most of the variability
properties of accreting systems
(\citealt{2001MNRAS.327..799K,2005MNRAS.359..345U,2006MNRAS.367..801A}),
including time lags.  Although hard lags are not the main subject of
this paper, we checked whether this model could explain the hard
positive lags in 1H0707-495. We followed the formulation of
\cite{2006MNRAS.367..801A} (equation 3 in particular), where the lag
depends on the emissivity of the disc, the fluctuations propagation
speed and accretion disc parameters (see
\citealt{2006MNRAS.367..801A} for details). The soft lags have to be taken as a separate component.
Fig. \ref{fig:fluc_model} shows the model from 
\cite{2006MNRAS.367..801A} along with the data. We used emission region size of few gravitational
radii and a steep emissivity profile as observed. The model does not represent a fit as
several combinations of parameters can reproduce the data, but it shows that it is consistent with
the observation. Although in the original model, the emission region is extended, lags for a
compact region can be produced if for example the Comptonising corona sees seed photons caused by
accretion flucutations before they reach and modulate the emission region, or if magnetic
energy is fed to the emission region from different radii. In fact, any model that is used to
explain the hard lags in GBH and AGN can be
used here, taking the soft lag as a separate component. In Fig. \ref{fig:fluc_model} we used a
simple uniform reflection transfer function of the form discussed in earlier.

\section{Discussion}\label{discussion}
There are fundamental differences between the small and large-scale
(inner and outer) reflector interepretations of the soft lag. We have
already shown that the small-scale reflection model of F09 and Z10
fits the energy spectrum, spectral variability and timing properties
of the data very well. This self-consistency is particularly apparent
in the striking match with the lag-energy spectrum in
Fig.~\ref{fig:lag_vs_energy}. A comparison of energy spectrum and
timing properties has not yet been performed for the large-scale outer
reflection model of M10b, because no self-consistent spectrum and
energy-dependent transfer-function has been produced for a large-scale
reflector. But it is clear that agreement between the spectrum and
timing data would need to satisfy stringent constraints.

The large-scale lag model has several problems. Any reflection model
predicts that there is a substantial contribution from scattered light
in the energy spectrum.  This would be seen as a blend of continuum
and reprocessed emission features mainly from Iron and
Oxygen. Absorption signatures are also expected to be imprinted in
such spectra (e.g. \citealt{2010MNRAS.404.1369S}).  Unless the
reflection is relativistically smeared (as in the case of the
small-scale inner reflector), these can be ruled out in 1H0707-495
because of the absence of any narrow emission or absorption features
in the spectrum (Z10, \citealt{2009MNRAS.399L.169B}).

Considering the simplicity of the models,  M10b suggest that the large-scale reflector
model is more parsimonious than that of the small-scale reflector, since in the small-scale
reflector model, soft and hard lags are attributed to two separate
processes, whereas the large-scale reflector can also produce soft lags. However, we have
demonstrated that in order to reproduce soft lags over a broad frequency range as observed, the
M10b model still requires a different transfer function in the soft band that is much narrower
compared to that in the hard band. Therefore, mathematically the two models are identical: both
require two different transfer function components to reproduce the lag-vs-frequency dependence. The
difference is that in the F09/Z10 model, the soft transfer function is naturally explained by highly
relativistic emission (iron L line in the soft band) from a small region. In the M10b model, it is
introduced to broaden the soft lags. Its physical origin has to be a small-scale reflector or one
that is close to the line of sight.

\begin{figure}
\centering
 \includegraphics[height=200pt,clip ]{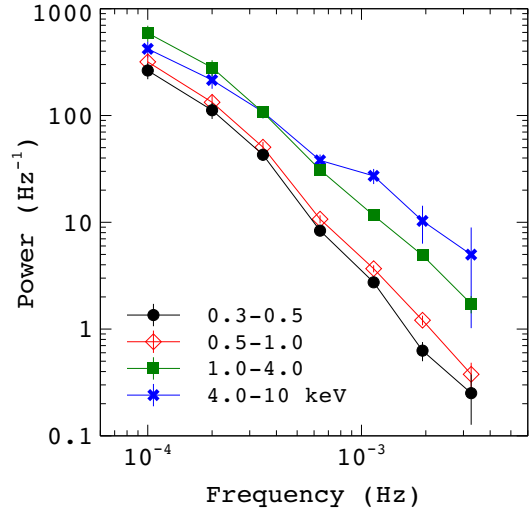}
\caption{The power spectra at different energies. Poisson noise has been subtracted by  fitting a
power-law + constant to the PSD and then subtracting the constant.}
\label{fig:psd_energy}
\end{figure}

Now we consider the consistency of the models with other observations.  If the variability is
dominated by two different processes, as suggested by the small-scale reflection model, one might
expect other timing properties to show evidence for this.  In fact, this appears to be the case.  In
Fig.~\ref{fig:psd_energy}, we show the power spectral densities of variations of 1H0707-495 in
several energy bands.  The hardest band shows extra power at high frequencies, which can be
explained if there is a separate high-frequency component of variability which shows a
different dependence on energy than the low-frequency component.  Similar behaviour has been seen in
the NLS~1 Ark~564, also corresponding to frequencies where there is a sharp drop to negative lags
\citep{2007MNRAS.382..985M}.  

Further evidence for a two-component
interpretation of the variability can be found in the coherence function (Fig. 18 in Z10). The
coherence is high at low and high frequencies ($\sim 0.9$, where 0 means the signals are not
coherent and 1 means they are fully coherent) and drops in between ($\sim 0.75$) in the overlap
which coincides with the transition from hard to soft lags.  The drop in coherence can be associated
with the transition between two separate variability components: where only one dominates the
coherence is maximised.  It is natural to associate the change in lag and coherence with the change
from a variability component dominated by propagation lags, to a faster component dominated by
light-travel effects, consistent with the small-scale reflection model.  Large-scale reflection
should produce the opposite effect in the PSD, smearing out variability in the hardest band relative
to the softer bands, contrary to what is observed, and would produce no discernable effect on the
coherence, since the same process is driving the variability at all frequencies.

Finally, we consider the physical plausibility of the small and large-scale reflection models for
explaining the observed time lags.  Crucially, both models require a soft reflection component
covering a narrow range of small lags relative to the direct continuum. The lags represent the
difference in path-lengths taken to the observer by light from the direct continuum, and from the
continuum source via the reflector.  For the disk-like geometry in the small-scale reflection model,
the small lags are then indicative of the small size of the reflector (which is the inner disk
within a few gravitational radii of the black hole) and its proximity to the continuum source. 
However, in the large-scale reflection model, the small lags do not closely correspond to the light
travel-time from the continuum to the reflector, because the reflector may have a non-disk-like
geometry, and the difference in path-lengths between direct and reflected light is strongly
dependent on that geometry.  In this way, the distance from the continuum to the reflector may be
substantially larger than the light-travel time. However, as we pointed out in
Section~\ref{dist_ref}, the distant reflector model remains a mathematical description of the lag
spectrum, with no self consistent physical picture. Furthermore, the model can be ruled out when
hard lags in Galactic black holes are considered. \cite{2001MNRAS.327..799K} have shown that the
predicted signatures of large-scale reflectors are not seen in the lag-energy spectra of Galactic
black holes, and yet hard lags are still seen, favouring a propagation interpretation for these
lags.

In conclusion, our latest timing results confirm the inner reflector
model, which also gives a self-consistent spectral model. The X-ray spectra
and timing of 1H0707-495 are dominated by emission and reflection from just
a few gravitational radii around the central black hole.

\section{Summary}
\begin{itemize}
\item The detailed lag-energy spectrum can be well modelled self-consistently with small-scale inner
reflection very close to the black hole. The contrary conclusions of
M10b are a consequence of not accounting for the blackbody component
dominating the spectrum at very soft energies.
\item Time delays, power spectra and the coherence functions all indicate the presence of two
variability processes: a low frequency component possibly due to propagating fluctuations, and a
high frequency small-scale reflector. These cannot be explained self-consistently with a
large-scale reflector, and in some cases, they are opposite to what it predicts.
\item The large-scale reflection model of M10b cannot produce soft lags over a broad frequency range
as observed without invoking an extra small-scale soft transfer function, due to clouds close to
the line of sight. This require a specific geomety putting the observer in a special position and
making it difficult to produce the observed spectrum.
\item The M10b lag model looks at lags independently of the energy spectrum. The lag spectrum puts
tight constraints on the ionisation of the distant reflector, where it has to be stronger in the
hard band. The model then predicts sharp emission and absorption features which are not observed.
\end{itemize}

\section*{Acknowledgements}
AZ thanks the Cambridge Overseas Trust and STFC. 
PU is supported by an STFC Advanced Fellowship and funding from the European Community's Seventh
Framework Programme (FP7/2007-2013) under grant agreement number ITN 215215 "Black Hole Universe".
ACF thanks the Royal Society for support. The authors thank the anonymous referee for their
comments.

\bibliography{bibliography}

\begin{thebibliography}{}

\bibitem[\protect\citeauthoryear{{Ar{\'e}valo}, {McHardy} \&
  {Summons}}{{Ar{\'e}valo} et~al.}{2008}]{2008MNRAS.388..211A}
{Ar{\'e}valo} P.,  {McHardy} I.~M.,    {Summons} D.~P.,  2008, MNRAS, 388, 211

\bibitem[\protect\citeauthoryear{{Ar{\'e}valo} \& {Uttley}}{{Ar{\'e}valo} \&
  {Uttley}}{2006}]{2006MNRAS.367..801A}
{Ar{\'e}valo} P.,  {Uttley} P.,  2006, MNRAS, 367, 801

\bibitem[\protect\citeauthoryear{{Blustin} \& {Fabian}}{{Blustin} \&
  {Fabian}}{2009}]{2009MNRAS.399L.169B}
{Blustin} A.~J.,  {Fabian} A.~C.,  2009, MNRAS, 399, L169

\bibitem[\protect\citeauthoryear{{Cackett}, {Miller}, {Bhattacharyya},
  {Grindlay}, {Homan}, {van der Klis}, {Miller}, {Strohmayer} \&
  {Wijnands}}{{Cackett} et~al.}{2008}]{2008ApJ...674..415C}
{Cackett} E.~M.,  {Miller} J.~M.,  {Bhattacharyya} S.,  {Grindlay} J.~E.,
  {Homan} J.,  {van der Klis} M.,  {Miller} M.~C.,  {Strohmayer} T.~E.,
  {Wijnands} R.,  2008, ApJ, 674, 415

\bibitem[\protect\citeauthoryear{{Fabian}, {Vaughan}, {Nandra}, {Iwasawa},
  {Ballantyne}, {Lee}, {De Rosa}, {Turner} \& {Young}}{{Fabian}
  et~al.}{2002}]{2002MNRAS.335L...1F}
{Fabian} A.~C.,  {Vaughan} S.,  {Nandra} K.,  {Iwasawa} K.,  {Ballantyne}
  D.~R.,  {Lee} J.~C.,  {De Rosa} A.,  {Turner} A.,    {Young} A.~J.,  2002,
  MNRAS, 335, L1

\bibitem[\protect\citeauthoryear{{Fabian}, {Zoghbi}, {Ross}, {Uttley}, {Gallo},
  {Brandt}, {Blustin}, {Boller}, {Caballero-Garcia}, {Larsson}, {Miller},
  {Miniutti}, {Ponti}, {Reis}, {Reynolds}, {Tanaka} \& {Young}}{{Fabian}
  et~al.}{2009}]{2009Natur.459..540F}
{Fabian} A.~C.,  {Zoghbi} A.,  {Ross} R.~R.,  {Uttley} P.,  {Gallo} L.~C.,
  {Brandt} W.~N.,  {Blustin} A.~J.,  {Boller} T.,  {Caballero-Garcia} M.~D.,
  {Larsson} J.,  {Miller} J.~M.,  {Miniutti} G.,  {Ponti} G.,  {Reis} R.~C.,
  {Reynolds} C.~S.,  {Tanaka} Y.,    {Young} A.~J.,  2009, Nature, 459, 540

\bibitem[\protect\citeauthoryear{{Kotov}, {Churazov} \& {Gilfanov}}{{Kotov}
  et~al.}{2001}]{2001MNRAS.327..799K}
{Kotov} O.,  {Churazov} E.,    {Gilfanov} M.,  2001, MNRAS, 327, 799

\bibitem[\protect\citeauthoryear{{Lyubarskii}}{{Lyubarskii}}{1997}]{1997MNRAS.%
292..679L}
{Lyubarskii} Y.~E.,  1997, MNRAS, 292, 679

\bibitem[\protect\citeauthoryear{{McHardy}, {Ar{\'e}valo}, {Uttley},
  {Papadakis}, {Summons}, {Brinkmann} \& {Page}}{{McHardy}
  et~al.}{2007}]{2007MNRAS.382..985M}
{McHardy} I.~M.,  {Ar{\'e}valo} P.,  {Uttley} P.,  {Papadakis} I.~E.,
  {Summons} D.~P.,  {Brinkmann} W.,    {Page} M.~J.,  2007, MNRAS, 382, 985

\bibitem[\protect\citeauthoryear{{McHardy}, {Papadakis}, {Uttley}, {Page} \&
  {Mason}}{{McHardy} et~al.}{2004}]{2004MNRAS.348..783M}
{McHardy} I.~M.,  {Papadakis} I.~E.,  {Uttley} P.,  {Page} M.~J.,    {Mason}
  K.~O.,  2004, MNRAS, 348, 783

\bibitem[\protect\citeauthoryear{{Miller}}{{Miller}}{2007}]{2007ARA&A..45..441%
M}
{Miller} J.~M.,  2007, ARA\&A, 45, 441

\bibitem[\protect\citeauthoryear{{Miller}, {Turner}, {Reeves} \&
  {Braito}}{{Miller} et~al.}{2010}]{2010arXiv1006.5035M}
{Miller} L.,  {Turner} T.~J.,  {Reeves} J.~N.,    {Braito} V.,  2010, ArXiv
  e-prints

\bibitem[\protect\citeauthoryear{{Miller}, {Turner}, {Reeves}, {Lobban},
  {Kraemer} \& {Crenshaw}}{{Miller} et~al.}{2010}]{2010MNRAS.403..196M}
{Miller} L.,  {Turner} T.~J.,  {Reeves} J.~N.,  {Lobban} A.,  {Kraemer} S.~B.,
    {Crenshaw} D.~M.,  2010, MNRAS, 403, 196

\bibitem[\protect\citeauthoryear{{Miyamoto} \& {Kitamoto}}{{Miyamoto} \&
  {Kitamoto}}{1989}]{1989Natur.342..773M}
{Miyamoto} S.,  {Kitamoto} S.,  1989, Nature, 342, 773

\bibitem[\protect\citeauthoryear{{Nandra}, {O'Neill}, {George} \&
  {Reeves}}{{Nandra} et~al.}{2007}]{2007MNRAS.382..194N}
{Nandra} K.,  {O'Neill} P.~M.,  {George} I.~M.,    {Reeves} J.~N.,  2007,
  MNRAS, 382, 194

\bibitem[\protect\citeauthoryear{{Nowak}, {Vaughan}, {Wilms}, {Dove} \&
  {Begelman}}{{Nowak} et~al.}{1999}]{1999ApJ...510..874N}
{Nowak} M.~A.,  {Vaughan} B.~A.,  {Wilms} J.,  {Dove} J.~B.,    {Begelman}
  M.~C.,  1999, ApJ, 510, 874

\bibitem[\protect\citeauthoryear{{Nowak}, {Wilms} \& {Dove}}{{Nowak}
  et~al.}{1999}]{1999ApJ...517..355N}
{Nowak} M.~A.,  {Wilms} J.,    {Dove} J.~B.,  1999, ApJ, 517, 355

\bibitem[\protect\citeauthoryear{{Pottschmidt}, {Wilms}, {Nowak}, {Heindl},
  {Smith} \& {Staubert}}{{Pottschmidt} et~al.}{2000}]{2000A&A...357L..17P}
{Pottschmidt} K.,  {Wilms} J.,  {Nowak} M.~A.,  {Heindl} W.~A.,  {Smith} D.~M.,
     {Staubert} R.,  2000, A\&A, 357, L17

\bibitem[\protect\citeauthoryear{{Poutanen}}{{Poutanen}}{2001}]{2001AdSpR..28.%
.267P}
{Poutanen} J.,  2001, Advances in Space Research, 28, 267

\bibitem[\protect\citeauthoryear{{Reis}, {Fabian} \& {Miller}}{{Reis}
  et~al.}{2010}]{2010MNRAS.402..836R}
{Reis} R.~C.,  {Fabian} A.~C.,    {Miller} J.~M.,  2010, MNRAS, 402, 836

\bibitem[\protect\citeauthoryear{{Revnivtsev}, {Gilfanov} \&
  {Churazov}}{{Revnivtsev} et~al.}{2000}]{2000A&A...363.1013R}
{Revnivtsev} M.,  {Gilfanov} M.,    {Churazov} E.,  2000, A\&A, 363, 1013

\bibitem[\protect\citeauthoryear{{Sim}, {Miller}, {Long}, {Turner} \&
  {Reeves}}{{Sim} et~al.}{2010}]{2010MNRAS.404.1369S}
{Sim} S.~A.,  {Miller} L.,  {Long} K.~S.,  {Turner} T.~J.,    {Reeves} J.~N.,
  2010, MNRAS, 404, 1369

\bibitem[\protect\citeauthoryear{{Tanaka}, {Nandra}, {Fabian}, {Inoue},
  {Otani}, {Dotani}, {Hayashida}, {Iwasawa}, {Kii}, {Kunieda}, {Makino} \&
  {Matsuoka}}{{Tanaka} et~al.}{1995}]{1995Natur.375..659T}
{Tanaka} Y.,  {Nandra} K.,  {Fabian} A.~C.,  {Inoue} H.,  {Otani} C.,  {Dotani}
  T.,  {Hayashida} K.,  {Iwasawa} K.,  {Kii} T.,  {Kunieda} H.,  {Makino} F.,
   {Matsuoka} M.,  1995, Nature, 375, 659

\bibitem[\protect\citeauthoryear{{Uttley}, {McHardy} \& {Vaughan}}{{Uttley}
  et~al.}{2005}]{2005MNRAS.359..345U}
{Uttley} P.,  {McHardy} I.~M.,    {Vaughan} S.,  2005, MNRAS, 359, 345

\bibitem[\protect\citeauthoryear{{Zoghbi}, {Fabian}, {Uttley}, {Miniutti},
  {Gallo}, {Reynolds}, {Miller} \& {Ponti}}{{Zoghbi}
  et~al.}{2010}]{2010MNRAS.401.2419Z}
{Zoghbi} A.,  {Fabian} A.~C.,  {Uttley} P.,  {Miniutti} G.,  {Gallo} L.~C.,
  {Reynolds} C.~S.,  {Miller} J.~M.,    {Ponti} G.,  2010, MNRAS, 401, 2419

\end{thebibliography}
\label{lastpage}

\end{document}